\documentstyle[twocolumn,aps]{revtex}
\def\c#1{\setbox0=\hbox{#1}\ifdim\ht0=1ex\accent24 #1%
  \else{\ooalign{\hidewidth\char24\hidewidth\crcr\unhbox0}}\fi}
\input epsf
\begin{document}
\draft
\twocolumn[\hsize\textwidth\columnwidth\hsize\csname@twocolumnfalse\endcsname

\title{Taxonomy of Stock Market Indices}
 
\author{Giovanni Bonanno$^{*}$, Nicolas Vandewalle $^{\dag}$ and 
Rosario N. Mantegna $^{*}$}

\address{
$^{*}$Istituto Nazionale per la Fisica della Materia, Unit\'a di Palermo, 
and \\ Dipartimento di Fisica e Tecnologie Relative, 
Universit\'a di Palermo, Viale delle Scienze, 
I-90128 Palermo, Italia\\ $^{\dag}$ Institut de Physique B5, Sart Tilman,
Universit\'{e} de Li\`{e}ge, \\ B-4000 Li\`{e}ge, Belgium.}

 
\maketitle
 
\begin{abstract}
We investigate sets of financial non-redundant and nonsynchronously
recorded time series. The sets are composed by a number of stock market 
indices located all over the world in five continents. By properly 
selecting the time horizon of returns and by using a reference 
currency we find a meaningful taxonomy. The detection of such a 
taxonomy proves that interpretable information can be stored in 
a set of nonsynchronously recorded time series.
\end{abstract}
\vskip1pc]

\narrowtext
 
One key aspect of information theory \cite{Shannon48} is that
unpredictable time series, namely time series which are not
or poorly redundant are characterized by statistical properties
which are almost indistinguishable from the ones observed
in basic random processes such as, for examples, Bernoulli
or Markov processes.
Within this theoretical framework it may appear paradoxical that
some time series generated in complex systems, which are playing a
vital role in biological and economic systems are essentially 
unpredictable and characterized by a negligible or pretty low
redundancy. Prominent examples are the time series of the price
changes of assets traded in a financial markets 
\cite{Cootner64,Campbell97,MS99,BP2000} 
and the symbolic series of coding regions of DNA 
\cite{Gatlin66,Mantegna9495}.

In this letter we show that an approximately non-redundant time 
series non-synchronously recorded may carry different levels 
of interpretable information provided that it can be analyzed 
synchronously together with other time series of the same kind. 
In other words we show that in addition to the information related to the
redundant nature of the time series other sources of information
may be present in a non-redundant time series and that such additional
information can be extracted by comparing the considered time 
series with analogous ones. Our work focuses on time series
monitoring financial markets located all over the world. 
With our study, we aim to detect in a quantitative way the existence
of links between different stock markets.

It is worth pointing out that
the study of the dynamics of stock exchange indices located 
all over the world has additional levels of complexity with respect to,
for example, the dynamics of a portfolio of stocks traded in 
a single stock market. To cite just two of the most prominent
ones -- (i) stock markets located all over the world have
different opening and closing hours; and (ii) transactions
in different markets are done by using different currencies 
that fluctuates themselves the one with respect to the other.
It is then important to quantify the degree of similarity 
between the dynamics of stock indices of nonsynchronous markets 
trading in different currencies. 

Here we present a study showing that meaningful
information can be extracted by a set of stock indices time 
series. In our study, the different levels of interdependence 
and complexity of data are elucidated by considering multiple 
applications of the same methodology on modified sets of the 
investigated time series.
In our study we are able to show that it is possible to extract 
a group of taxonomies that directly reflects geographical 
and economic links between several countries all over the years.
This is obtained by using the almost non-redundant time 
series of several stock indices of financial markets located  
all over the world only.

The efficient market paradigm states that stock returns of financial 
price time series are unpredictable \cite{Samuelson65}. Within this 
paradigm, time evolution of stock returns is well described by a 
random process \cite{Cootner64}. Several empirical analyses of real 
market data have proven that returns time series are approximately 
described by unpredictable non-redundant time series 
\cite{Cootner64,MS99,BP2000}.
The absence of redundancy is not complete in real markets and the 
presence of residual redundancy has been detected 
\cite{Campbell97,Baviera99}. A minimized degree of redundancy is 
required to avoid the presence of arbitrage opportunities.
  
We investigate two sets of data -- (i) the nonsynchronous time 
evolution of $n=24$ daily stock market indices computed in local 
currencies during the time period from January 1988 to December 1996,
and (ii) the closure value of the 51 Morgan Stanley Capital 
International (MSCI) country indices
daily computed in local currencies or in USA dollars in the
time period from January 1996 to December 1999.   
The stock indices used in our research belong to stock markets 
distributed all over the world in five continents.

We already stated that 
a set of stock indices time series is essentially different
from a portfolio of stocks traded in a single
stock market. Specifically, the fact that trading may occur at 
different time in two different cities implies that some 
markets are open during the time whereas others are closed 
(the most prominent example concerns New York and Tokyo 
stock markets). This makes impossible a rigorously 
synchronous analysis of a large number of stock indices 
located all over the world. An 
analysis of daily data of say closure values may induce spurious 
correlations introduced just by the specific time at which
the variables are stored. The effects of nonsynchronous
trading in time series analysis is well documented in the 
economic literature \cite{Lo90,Becker90,Lin94}. In fact
different degrees of correlation between the New York and
Tokyo markets are estimated depending if one consider
the closure - closure between the two markets or the
closure - opening. In particular it has been empirically
detected that the highest degree of correlation between 
these two markets is observed between the open-closure 
return of the New York stock exchange at day $t$ and 
the opening-closure of the
Tokyo stock market at day $t+1$ \cite{Becker90}.

The aim of this study is to consider a large set of indices. 
It is of course impossible
to collect a set of indices located all over the world 
which are synchronous with respect to the opening
and closing hours. This intrinsic limitation motivate us
to consider a week time horizon where the nonsynchronous 
hourly mismatch of our data is minimized.

We aim to discover the presence of {\it interpretable information} 
in a set of time series. We proceed by determining   
a quasi-synchronous correlation coefficient of the
weekly difference of logarithm of closure value 
of indices. The correlation coefficient is 
\begin{equation}
\rho_{ij}=\frac{<Y_i Y_j>-<Y_i><Y_j>}
{\sqrt{(<Y_i^2>-<Y_i>^2)(<Y_j^2>-<Y_j>^2)}}
\end{equation}
where $i$ and $j$ are the numerical labels of indices, 
$Y_i=\ln S_i(t)-\ln S_i(t-1)$ and $S_i(t)$ is the 
last value of the trading week $t$ for the index $i$. 
The correlation coefficient is 
computed between all the possible pairs of indices present in 
the database. The statistical average is a temporal average 
performed on all the trading days of the investigated time period.
We then obtain the $n \times n$ matrix of 
correlation coefficient for weekly logarithm index differences 
(which almost coincides with index returns). 
Correlation matrices have been recently investigated within
the framework of random matrix theory \cite{Laloux99,Plerou99}. 
Here we take a different perspective, 
we use the method introduced in ref. \cite{Mantegna99}.
Specifically we assume that the subdominant ultrametric space 
associated to a metric distance may reveal part of the 
economic information stored in the time series.
This is obtained by defining
a quantitative distance between each pair of elements $i$ and
$j$, $d(i,j)=\sqrt{2(1-\rho_{ij})}$ and then using this 
distance matrix $\bf D$ to determine the minimum spanning 
tree (MST) connecting 
the $n$ indices. 
The MST, a theoretical concept of graph theory 
\cite{West96}, allows to obtain, 
in a direct and unique way, the subdominant ultrametric space 
and the hierarchical organization of the elements 
(indices in our case) of the investigated data set.
Subdominant ultrametric space \cite{Rammal86} has been 
fruitfully used in the description of frustrated complex systems. 
The archetype of this kind of systems is a spin glass \cite{Mezard87}. 

In the rest of this letter, we show that the group 
of taxonomies found by considering the subdominant ultrametric 
matrices  $\bf{D^<}$ 
associated with the distance matrices $\bf{D}$, obtained from 
different sets of quasi-synchronously time series investigated 
in local currencies or in USA Dollars, are of direct interpretation. 

We first investigated the set of 24 indices of 20 different 
countries recorded during the period 1988-1996. We divide 
the entire period in 6 four years partially overlapping periods. The 
first covers the years 1988-1991, the second 1989-1992 and
so on. Each 4 years period comprises 207 or 208 week records for
each time series. In all the periods we detect distinct
clusters of North-America, Europe and
Asia-Pacific stock indices. The North-America cluster 
is rather stable over the years and includes the USA indices 
Dow Jones 30, Standard \& Poor's 500, Nasdaq 100 and Nasdaq 
Composite. The European cluster increases
in size starting in the first period as the one formed 
by Amsterdam AEX, Paris CAC40, Frankfurt DAX and
London FTSE and ending as a FTSE, AEX, DAX, CAC40, Madrid General and
Oslo General cluster in the last period. Milan Comit index stays 
always out of the European cluster in the investigated periods. 
This is not so surprising because Italy was the 
only large European economy rather far from the so-called 
Maastricht parameters during that period. 
The Asian-Pacific cluster is also 
expanding as time goes on. It starts as a Kuala Lumpur Comp., 
Singapore Straits Times Industrial and Bangkok SET cluster and ends
as a Kuala Lumpur, Singapore, Hong Kong Hang-Seng, Bangkok, 
Australia All Ordinary, Jakarta Comp. and Philippines Comp.
cluster. Japanese stock indices do not join the Asian-Pacific cluster 
and Japan behaves as a poorly linked country. The same occurs 
for BSE30 index (of India) and South-America indices. 

In Fig. 1 we show the hierarchical trees obtained for the
first and the last averaging time period. The presence 
of clusters is observed in both periods but the tree
of the second period has larger clusters. 
In summary our study shows that regional links between 
different economies emerge directly from time series. 
Moreover, an increase of the size of observed clusters 
and a relative stability of the clusters  
over the years is detected.

With the aim of expanding this analysis over one of the 
largest sets of indices today available, we consider the 
set of 51 world indices computed by MSCI. For a so large 
set of indices the point of view
of the investor becomes crucial. In other words it is
important to consider the problem also from the
perspective of an international investor
simultaneously monitoring the various markets.
Several aspects
of the different countries needed to be taken into
account to make an appropriate comparison, they include
the difference in currency values, levels of taxation etc..
Here we consider the most important of these
differences namely the fact that the performances of different
stock markets need to be
compared by an international investor by using one 
reference currency. To evaluate the impact of a
change of currency in the computation of indices, 
we consider the 51 MSCI country indices either in local
currencies and in USA Dollars.

The 51 indices belongs to 51 different countries located
in all continents. They comprises so-called emerged and
emerging markets. Indices can be found
at the web site http://www.mscidata.com. The data are 
daily data and covers the period 1996-1999. In Fig. 2a
we show the result of our analysis performed by investigating
weekly closure data in local currencies during the period 1996-1999.
Four distinct clusters are detected (indicated in the
bottom of the figure by a solid line). The cluster number one is 
essentially a North-American (green lines indicating USA and 
Canada indices) and European cluster
(blue lines).
There is only one country index from the Asia-Pacific area 
and it is Australia (red line). The cluster number
two comprises 4 South-America country indices and the number
three is composed by 6 Asia-Pacific country indices 
whereas the small cluster number four comprises India 
and Pakistan. The only world
region that does not explicitly show index clustering 
is the world region of Africa-Middle East (purple lines). 
However, it is worth noting
that several of these country indices are found at the extreme right
of the hierarchical tree namely they are all quite far from
any other country.
Once again Japan index is disconnected from the Asia-Pacific
cluster and is observed at the external edge of the 
South-America cluster. Between European countries the ones which
are outside cluster one are The Czech Republic, Greece, Turkey and
Luxembourg. Of these four countries only the Luxembourg is 
considered by MSCI an emerged market.

The same analysis is then repeated for the same indices in the 
same period but using indices computed in USA Dollars. 
The hierarchical tree of this investigation is shown in 
Fig, 2b. The overall structure observed in Fig. 2a is
conserved but some relevant changes are detected.
For example the Australian index leaves cluster one
and links together with New Zealand in cluster three
of this figure. Japan moves still far being now the first 
read line after the Asian-Pacific cluster, the small
India-Pakistan cluster disappears and Peru' links at the
edge of cluster one. In summary the results of our analysis 
show that the
computing of the indices in a single reference 
currency can modify the obtained hierarchical structure.
However, the changes detected in the specific investigated
period are not dramatic and limited to few countries.

To verify if the nonsynchronous recording of daily data
indeed affects our findings we also determine the hierarchical
tree for daily closure changes for the same set
of indices used to obtain the tree of Fig. 2b. 
This new hierarchical tree shows the same overall structure observed
in the tree of Fig. 2a but with a number of different links which
are probably induced by the use of nonsynchronous
time series. Specifically we observe that almost all
the American indices cluster together (Brazil, Argentina, Mexico, USA,
Canada and Peru') and South-Africa cluster with the 
(in this case just) European cluster.
      
In conclusion, we have shown that sets of stock index time series
located all over the world can be used to extract
economic information about the links between different
economies provided that the effects of the nonsynchronous
nature of the time series and of the different currencies
used to compute the indices are properly taken into account.

 
We are grateful to Fabrizio Lillo for fruitful discussions
and to an anonymous referee for helpful comments.
G. Bonanno and R.N. Mantegna wish to thank INFM, ASI and 
MURST for financial support, N. Vandewalle is financially 
supported by FNRS.

\begin{figure}
\epsfxsize=2.6in
\centering \epsfbox{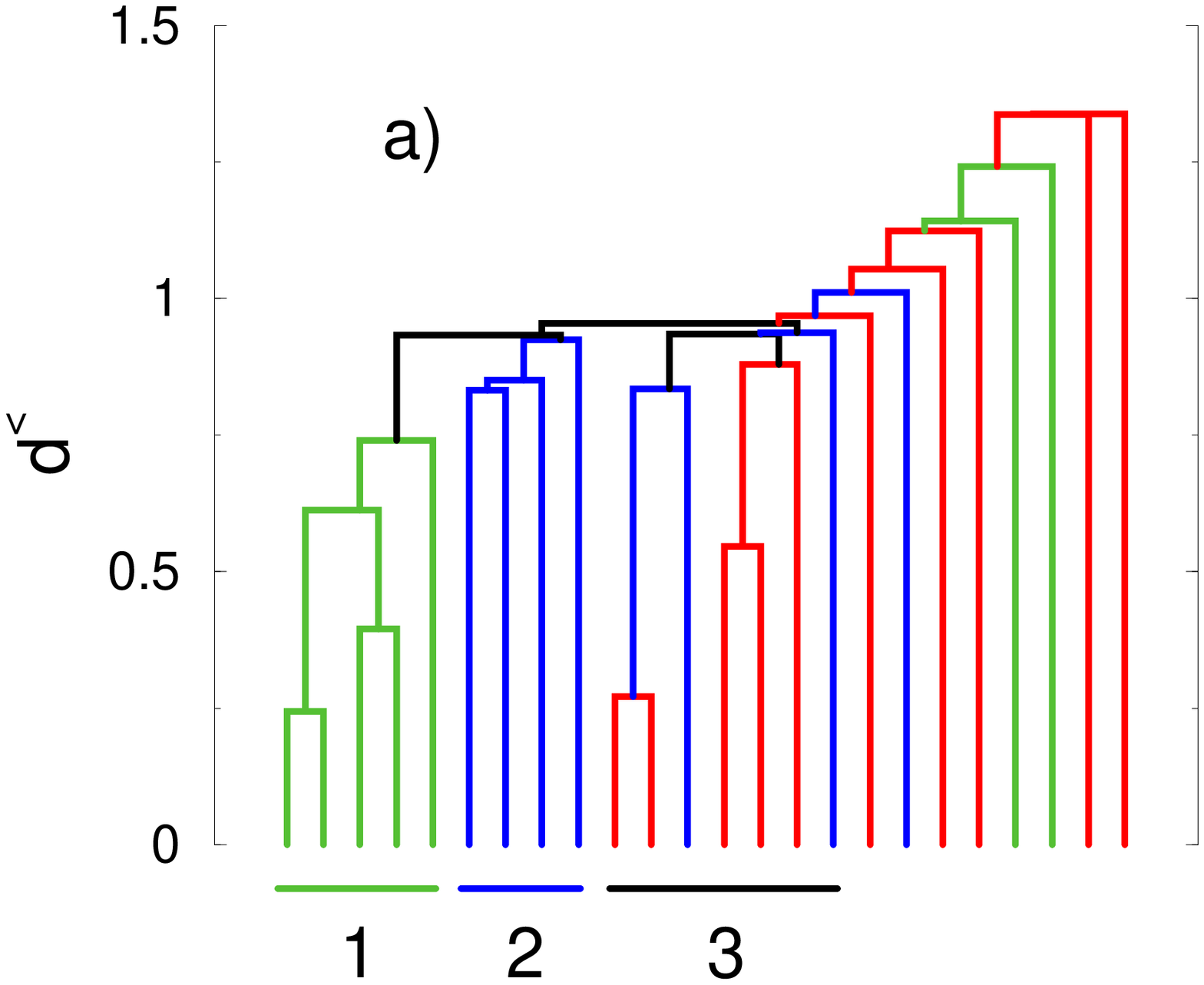}

\vspace{12pt}
\epsfxsize=2.6in
\centering \epsfbox{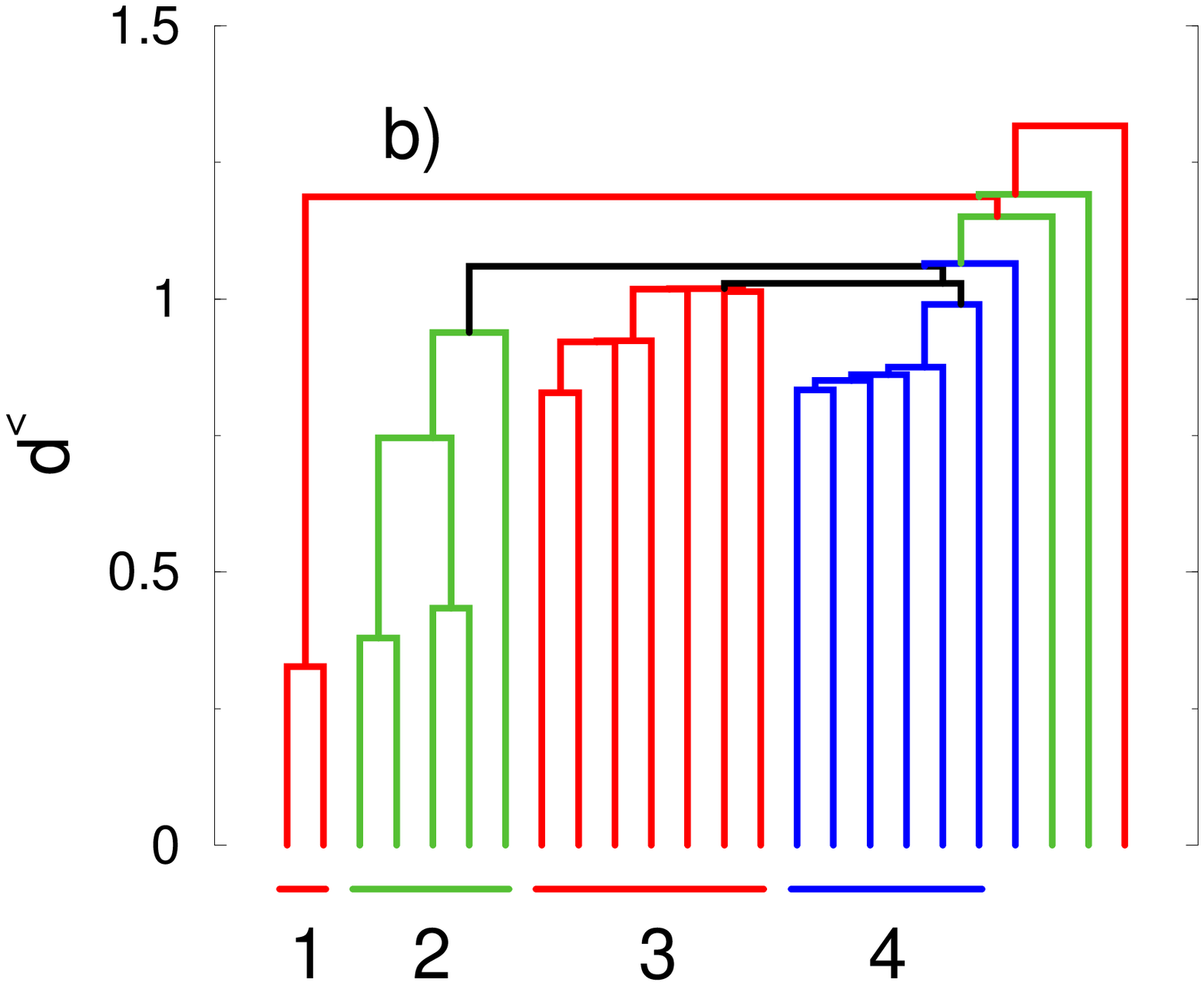}

\vspace{12pt}
\caption{(a) Hierarchical tree obtained starting 
from the correlation coefficients $\rho_{ij}$ of the 
set of 24 stock indices of weekly data. The correlation
coefficient is obtained by averaging the index changes
during the time period 1988 - 1991. Each line refers to an index. 
Colors of the index country are coded in the following 
way: Americas (green), Asia-Pacific (red) and Europe (blue).
The regional clusters of North-America (cluster number one, 
indices: USA DJIA, S\&P 500, NASDAQ Comp., Nasdaq 100 
and Canadian Toronto SE300) and Europe (cluster number two,
indices: Amsterdam, Dax, CAC40 and FTSE) are detected.
A third cluster is also present but it mixes indices
of different world regions (Indices: TOPIX, Nikkei 225, 
Madrid General, Kuala Lumpur, Singapore, Bangkok SET and Milan Comit). 
Remaining
indices are Hang Seng, Oslo, Australia All Ord., Philippines, 
Mexico IPC, Chile IGPA, Jakarta and Bombay BSE30 from left to
right respectively.
(b) As in a) but for the time period 1993-1996.
The clusters detected in this period are larger
and more homogeneous. Specifically, cluster one is 
a Japanese cluster (TOPIX and Nikkei 225 indices),
cluster number two contains North-America indices as in a),
cluster three is the Asia-Pacific cluster consisting of
Kuala Lumpur, Singapore, Hang Seng, Bangkok SET, Australia All Ord.,
Jakarta and Philippines indices whereas cluster four is the 
European cluster formed by FTSE, Amsterdam, DAX, CAC40, Madrid General
and Oslo. The remaining indices are Comit, Mexico IPC,
Chile IGPA and BSE30 from left to right respectively.}
\label{fig1}
\end{figure}
 
\begin{figure}[t]
\epsfxsize=2.6in
\centering \epsfbox{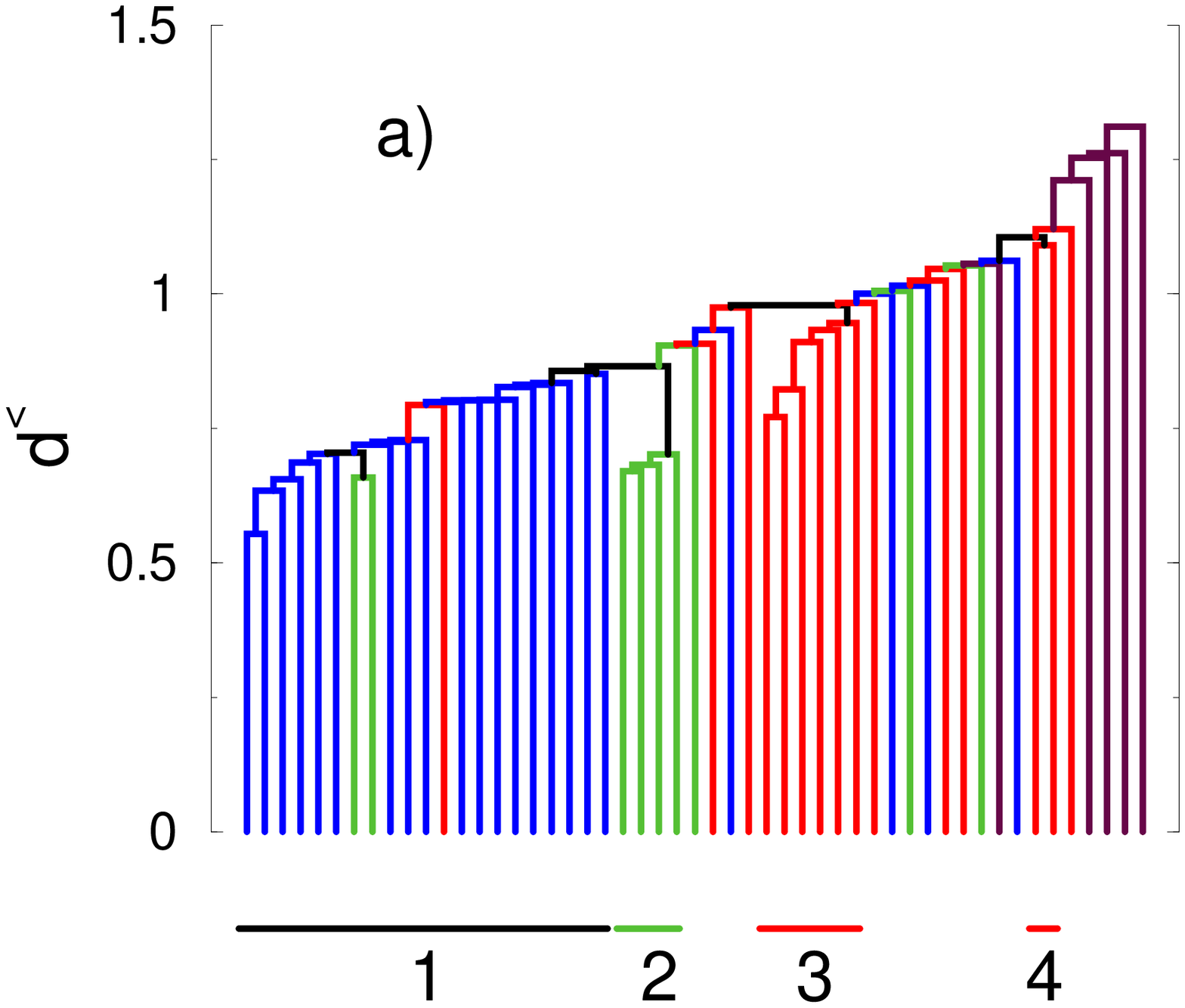}

\epsfxsize=2.6in
\centering \epsfbox{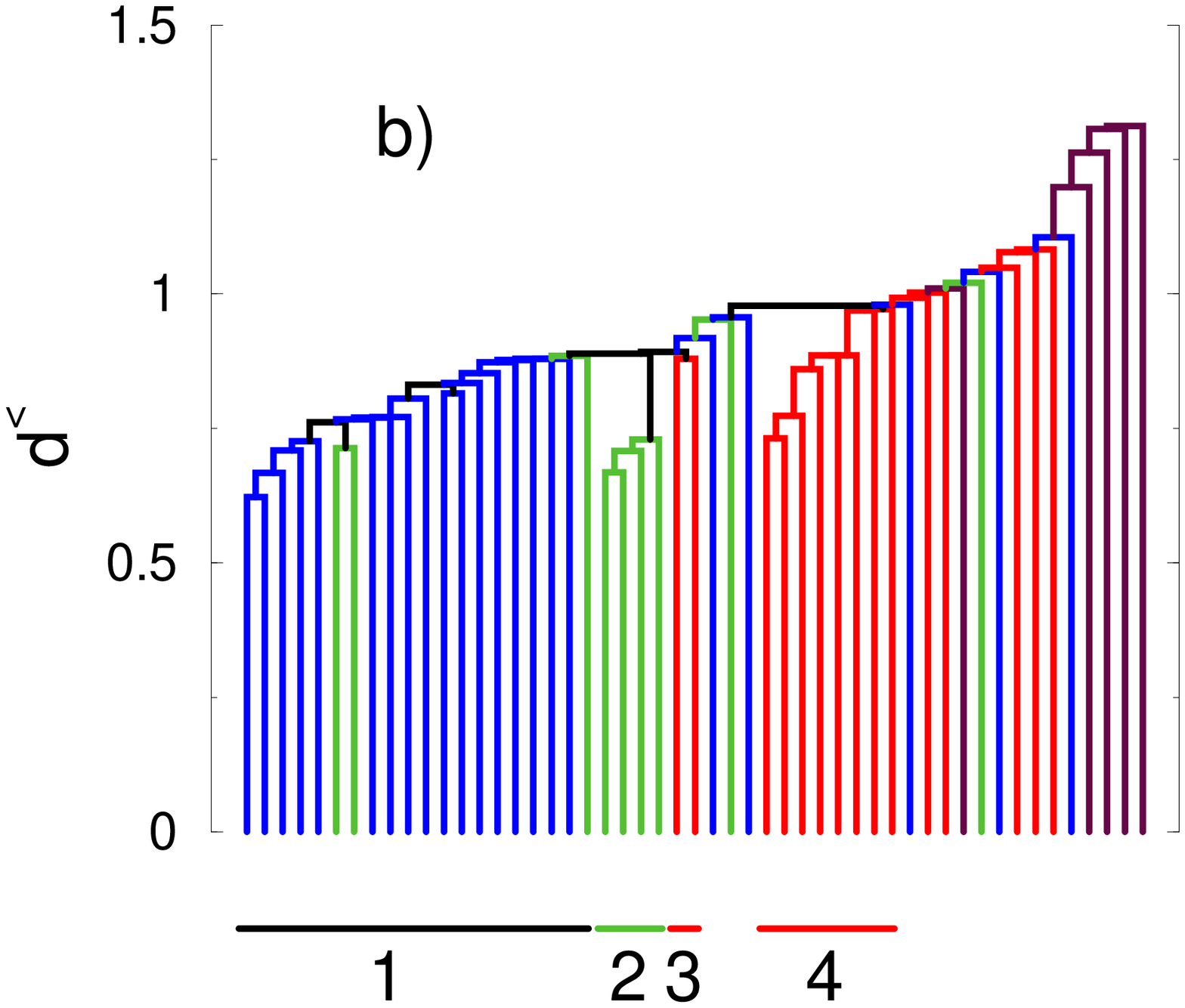}

\vspace{12pt}
\caption{(a) Hierarchical tree obtained starting 
from the correlation coefficients $\rho_{ij}$ of the 
set of 51 MSCI of weekly data computed in local currencies. 
The correlation coefficient is obtained by averaging 
the index changes during the time period 1996 - 1999. 
Each line refers to an index. 
Colors are coded
as in Fig. 1 but in addition Africa-Middle East countries are also
present and their color is purple.
Four clusters are observed. Cluster number one contains
France, Germany, Sweden, Netherlands, Switzerland,
Spain, USA, Canada, Great Britain, Finland, Italy,
Australia, Belgium, Norway, Austria, Denmark, Russia, Ireland,
Portugal, Poland and Hungary indices. It is a North-American and European
cluster with the addition of Australia index. Cluster two is composed
by Chile, Brazil, Argentina and Mexico indices they belong all 
to South-American countries.
Externally linked to the previous two clusters we find
Per\'u, Japan, The Czech Republic and New Zealand indices. 
Cluster three is an Asian-Pacific cluster where we find Singapore,
Hong Kong, Philippines, Malaysia, Thailand and China indices.
Just after this cluster we have Indonesia, Greece, Venezuela,
Turkey, Taiwan, Sri Lanka, Colombia, South-Africa and Luxembourg
indices. The last cluster
is the small cluster of India and Pakistan indices. The remaining
country indices are Korea, Egypt, Israel, Jordan and Morocco.
(b) As in a) but for indices computed in USA Dollars. The country 
indices are from left to right:
France, Germany, Sweden, Finland, Netherlands, 
USA, Canada, Spain, Great Britain, Norway, Italy, Poland, Hungary, 
Switzerland, Russia, Belgium,  Austria, Portugal, Denmark, Peru' 
(end of cluster one), Mexico, Argentina, Brazil, and Chile (end
of cluster two), New Zealand and Australia (cluster three),
Ireland, Venezuela, The Czech Republic, (start of cluster
four) Singapore,
Hong Kong, Philippines, Thailand, China and Indonesia (end of cluster four),
Taiwan, Malaysia, Greece, Japan, Korea, South-Africa, Colombia,
Turkey, Sri Lanka, India, Pakistan, Luxembourg, Egypt, Israel,
Jordan and Morocco.}   
\label{fig2}
\end{figure}


\begin{references}

\bibitem{Shannon48}
E.C. Shannon, {\it Bell System Tech. J.\/} {\bf 27}, 379 (1948).

\bibitem{Cootner64}
P.H. Cootner, ed., {\it The Random Character of Stock Market 
Prices\/} (MIT Press, Cambridge MA, 1964).

\bibitem{Campbell97}
J.Y. Campbell, A.W. Lo, and A.C. MacKinlay, {\it The Econometrics 
of Financial Markets\/} (Princeton University Press, Princeton, 1997).

\bibitem{MS99}
R.N. Mantegna and H.E. Stanley, {\it An Introduction to 
Econophysics: Correlations and Complexity in Finance\/} 
(Cambridge University Press, Cambridge, 2000).

\bibitem{BP2000}
J.-P. Bouchaud and M. Potters, {\it Theory of Financial Risks}
(Cambridge University Press, Cambridge, 2000).

\bibitem{Gatlin66}
L.L. Gatlin, {\it J. Theor. Biol.\/} {\bf 10}, 281 (1966).

\bibitem{Mantegna9495}
R.N. Mantegna, {\it et al.}, {\it Phys. Rev. Lett.\/} {\bf 73}
, 3169 (1994).
{\it Phys. Rev.\/} A{\bf 52}, 2939 (1995).

\bibitem{Samuelson65} P.A. Samuelson, {\it Industrial Management
Review\/} Vol. 6, No. 2 , 41-49 (1965).

\bibitem{Mantegna96}
R.N. Mantegna and H.E. Stanley, {\it Nature\/} {\bf 383}, 587 (1996).

\bibitem{Baviera99}
R. Baviera, {\it et al.}, 
`Efficiency in foreign exchange markets', Cond.-Mat. pre-print 
server 9901225.


\bibitem{Lo90} A.W. Lo and A.C. MacKinlay, {\it Journal of 
Econometrics} {\bf 45}, 181 (1990).

\bibitem{Becker90} K.G. Becker, J.E. Finnerty, and M. Gupta,
{\it Journal of Finance} {\bf XLV}, 1297 (1990).

\bibitem{Lin94} W.-L. Lin, R.F. Engle, T. Ito,
{\it The Review of Financial Studies} {\bf 7}, 507 (1994).

\bibitem{Laloux99}
L. Laloux, {\it et al.}, {\it Phys. Rev. Lett.\/} {\bf 83}, 
1467 (1999).

\bibitem{Plerou99}
V. Plerou, {\it et al.}, {\it Phys. Rev. Lett.\/} {\bf 83}, 
1471 (1999).

\bibitem{Mantegna99}
R.N. Mantegna, {\it Eur. Phys. J. B\/} {\bf 11}, 193 (1999).


\bibitem{West96}
D.B. West, {\it Introduction to Graph Theory\/} (Prentice-Hall,
Englewood Cliffs NJ, 1996).

\bibitem{Rammal86}
R. Rammal, G. Toulouse, and M.A. Virasoro, {\it Rev. Mod. Phys.} 
{\bf 58}, 765 (1986).

\bibitem{Mezard87}
M. M\'ezard, G. Parisi, and M.A. Virasoro,
{\it Spin Glass Theory and Beyond \/} (World Scientific, 
Singapore, 1987).

\end{references}
\end{document}